# Fast localization and single-pixel imaging of the moving object using time-division multiplexing

Zijun Guo,[1,2] Wenwen Meng,[3,4] Dongfeng Shi,[1,2,3,*] Linbin Zha,[1,2]  Wei Yang,[1,2] Jian Huang,[1,3] Yafeng Chen,[1] and Yingjian Wang[1,2,3]

[1]Key Laboratory of Atmospheric Optics, Anhui Institute of Optics and Fine Mechanics, Hefei Institutes of Physical Science, Chinese Academy of Sciences, Hefei 230031, China
[2]University of Science and Technology of China, Hefei 230026, China
[3]Advanced Laser Technology Laboratory of Anhui Province, Hefei 230037, China
[4]School of Artificial Intelligence and Big Data, Hefei University, Hefei 230601, China

*Corresponding author.
  E-mail address: dfshi@aiofm.ac.cn

**Abstract:** When imaging moving objects, single-pixel imaging produces motion blur. This paper proposes a new single-pixel imaging method, which can achieve anti-motion blur imaging of a fast-moving object. The geometric moment patterns and Hadamard patterns are used to alternately encode the position information and the image information of the object with time-division multiplexing. In the reconstruction process, the object position information is extracted independently and combining motion-compensation reconstruction algorithm to decouple the object motion from image information. As a result, the anti-motion blur image and the high frame rate object positions are obtained. Experimental results show that for a moving object with an angular velocity of up to 0.5rad/s relative to the imaging system, the proposed method achieves a localization frequency of 5.55kHz, and gradually reconstructs a clear image of the fast-moving object with a pseudo resolution of $512\times512$. The method has application prospects in single-pixel imaging of the fast-moving object.

## 1. Introduction

Single-pixel imaging (SPI) performs a series of modulations on the object, collects the reflected intensity from the object scene, and uses the reconstruction algorithm to obtain the object image. This imaging method sacrifices temporal resolution for spatial resolution. Its temporal resolution is low [1]. As a result, when imaging moving objects, the object's motion will be coupled to the image information, and the image will be blurred [2,3].

The traditional imaging system can shorten the single exposure time to make the moving object in an equivalent static state and record a clear object image. The existing high-speed camera can shorten the single exposure time to less than 1/10000 seconds [4,5]. It can photograph the bullet or the crack propagation at the moment of glass breakage. However, SPI is different from traditional imaging. The single-frame imaging time of SPI is difficult to shorten to this level while requiring high imaging quality. Although in recent years, researchers have done much work to improve the frame rate of single-pixel imaging, such as high-speed LED array as a structured light generation device [6], modulation based on orthogonal coding patterns, such as Fourier[7-12], Radon [13-14], compressed sensing algorithm [15-19], sorted Hadamard patterns in various ways, such as Walsh, Cake-Cut, Russian Doll [20-23]. Even though, there is still a big gap in single-frame imaging time between single-pixel imaging and traditional imaging, especially in the case of high-resolution imaging.

Because the single frame imaging time of SPI is difficult to compress, some researchers have adopted other methods to eliminate imaging motion blur. For example, Jiao et al. [24] uses a global optimization algorithm to estimate motion parameters and optimize the restored image in the reconstruction process, realizing the clear imaging of the moving object. Sun et al. [25] reconstructed images of the high-speed moving object using static pseudo-random patterns combined with the compressed sensing algorithm. However, prior knowledge of object motion types, such as rotation or translation, or velocity, is required in the above methods.

The above methods use prior parameters of object motion to rectify the reconstructed blur image. To get rid of the dependence on motion prior parameters, some researchers obtain object motion information from low-resolution images. Sun et al. [26] used the cross-correlation between a series of low-quality images to obtain the motion displacement of the object and then used the motion information to perform compensation, superimposing the low-quality images and gradually reconstructing a clear image of the object. However, at least 300 samples are required to obtain a low-quality image for localization. Wu et al. [27] used the Cake-Cut ordering Hadamard illumination patterns and periodically used the top-ranked patterns to acquire low-quality images quickly. Locate the moving object through those images to achieve anti-blur reconstruction. Image a slow-moving object and achieve a positioning frequency of 150fps. These methods obtain the target position through the correlation among the blurred images, and its positioning frequency is low. For example, Wu et al. use a relatively low imaging resolution of $64 \times 64$ pixels. At least 30 samplings are still required for one positioning [27]. The required sampling ratio is about 0.8%. The accuracy of positioning parameters is closely related to the number of samples. Improve positioning accuracy need to increase the number of samplings. That also leads to a longer sampling time and decreases the positioning frequency, which directly increases the interference caused by object motion and reduces the effect of anti-motion blur imaging.

The above two methods use the object's low-quality images to obtain motion information. There are mutual constraints on the positioning frequency and accuracy. Sun et al. [28] also proposed another method directly using the correlation between detected signals to locate moving objects. But the number of samples required for each positioning is still over a hundred. Taking the DMD modulator as an example, these methods [26-28] just can locate the moving object with hundreds of frames per second and achieve anti-motion blur imaging of the moving object at a relatively low speed. Our recently published paper [29] also proposes a different method that achieves 67.3 Hz positioning frequency and restores the moving object video with better quality.

In previous works, our group used a single-pixel detector to achieve real-time localization of the moving object up to 11.1kHz [30,31] without imaging. The advantage of this method is that the number of modulations and samples required for object positioning is just three. Moreover, this method preserves a relatively high positioning accuracy of 5.46 pixels with 768×768 resolution patterns. This feature provides the condition for positioning with both high accuracy and frequency. Based on this method, this paper uses the geometric moment patterns and Hadamard patterns to alternately encode the position information and the image information of the object with time-division multiplexing. In the reconstruction process, the object position information is extracted independently and combining motion-compensated reconstruction algorithms to decouple the object motion from image information. The anti-motion blur image could be constructed after the above procedure.

In the experiment, the 5.55kHz localization and anti-motion blur single-pixel imaging of a moving object with a pseudo resolution of 512×512 pixels was simultaneously realized. And combined with the obtained high frame rate object positions, the video of the moving object is obtained. This method has the prospect of expanding the application range of single-pixel imaging for fast-moving objects. The second part will introduce the method. The third part will use simulation and experiment to verify the proposed method's effectiveness. Finally, the method will be summarized and prospected.

## 2. Method

For conventional cameras, when imaging a moving object, keeping the object in a quasi-stationary state relative to the imaging system within the single-frame exposure time is necessary. It can be described as "frozen in time". Since it is difficult to shorten the single frame imaging time for the single-pixel imaging system, "frozen in time" cannot be effectively realized. However, the object can be kept quasi-stationary in another way by obtaining the displacement of the object and compensating in projection space, called "frozen in projection space". During the reconstruction process, the projection patterns are subjected to equal displacement compensation in the opposite direction of the object moves. That makes the moving object remains stationary relative to the projection space. For single-pixel imaging system, the acquisition process of a moving object can be expressed as follows:

$$B_i = \sum_{x,y} P(x+r_{x,i}, y+r_{y,i}) H_i(x, y) = \sum_{x,y} P(x, y) H_i(x-r_{x,i}, y-r_{y,i}), \tag{1}$$

where $B_i$ represents the detected bucket signals; $H_i$ is the $i$-th structured light to illuminate the object, it is based on the Hadamard matrix; $P$ is the reflectivity of the object, $x$ and $y$ represent the coordinates, $r_{x,i}$ and $r_{y,i}$ represent the displacement of the object relative to the initial position (assuming it is the center of the imaging view, as shown in Fig. 1. Position 0) within the measurement time (time since imaging/measurement started), expressed as:

$$r_{x,i} = \int_0^{t_i} v_x(t) dt \tag{2}$$

$$r_{y,i} = \int_0^{t_i} v_y(t) dt, \tag{3}$$

where $v_x$ and $v_y$ represent the object moving speed in the two directions. $t_i$ is the time from the measurement start to the projection of pattern $H_i$. Eq. (1) shows that "frozen in projection space" can be equivalently realized by reverse displacement of the structured illumination patterns, so the reconstruction with motion compensation of the image is:

$$P'(x, y) = \frac{1}{\eta} \sum_{i=1}^{\eta} B_i H_i(x-r_{x,i}, y-r_{y,i}), \tag{4}$$

where $\eta$ is the number of illumination patterns, $P'(x, y)$ is the reconstructed anti-motion blur image using the motion compensation single-pixel imaging (MCSPI). The above procedure is shown in Fig. 1. Based on the above analysis, to reconstruct the anti-motion blur image with motion compensation, it is necessary to obtain the displacement $r_x$ and $r_y$ of the object relative to the center of the view. The following text will describe how to obtain the displacements.

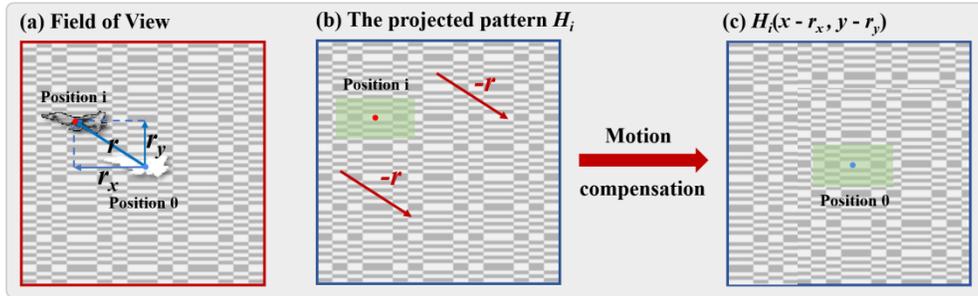

**Fig. 1.** The principle of the motion compensation reconstruction: (a) the object moving in the field of view. (b) shows the projected pattern $H_i$, **-r** is the vector by which $H_i$ is shifted. (c) $H_i(x-r_x, y-r_y)$ is used for reconstruction.

In order to obtain the displacements of the moving object, the single-pixel moment detection method [30,31] is used to locate the moving object in real-time. For the image $P(x,y)$, its $u+v$ order geometric moment can be calculated by the following formula:

$$m_{u,v} = \sum_{x=1}^{C} \sum_{y=1}^{R} x^u y^v P(x, y), \quad u, v = 0, 1, 2... \tag{5}$$

where $C$ and $R$ represent the number of columns and rows of the image. For a grayscale image, the zero-order moment $m_{00}$ represents the accumulation of the image grayscale. The first-order moments $m_{01}$ and $m_{10}$ describe the moment of the image about the $x$-axis and $y$-axis, respectively, which can be used to calculate the image centroid and reflect the object position. The centroids $x_c$ and $y_c$ can be expressed as follows:

$$x_c = m_{10} / m_{00} , \quad y_c = m_{01} / m_{00}. \tag{6}$$

In the SPI system, after illuminating the object with the patterns, the detected signal reflects the correlation of the object and the modulation patterns, which is mathematically described as:

$$I_k = \sum_{x,y} P(x,y) S_k(x,y), \tag{7}$$

where matrices $S_k$ are used, which are:

$$S_1 = \begin{bmatrix} N & N & \cdots & N \\ N & N & \cdots & N \\ \vdots & \vdots & \ddots & \vdots \\ N & N & \cdots & N \end{bmatrix}, S_2 = \begin{bmatrix} 1 & 2 & \cdots & N \\ 1 & 2 & \cdots & N \\ \vdots & \vdots & \ddots & \vdots \\ 1 & 2 & \cdots & N \end{bmatrix}, S_3 = \begin{bmatrix} N & N & \cdots & N \\ N-1 & N-1 & \cdots & N-1 \\ \vdots & \vdots & \ddots & \vdots \\ 1 & 1 & \cdots & 1 \end{bmatrix}. \tag{8}$$

The matrices $S_1$, $S_2$, $S_3$ are derived from Eq.5. These matrices are artificially set to obtain geometric moment values using the SPI system. These matrices are used to interact with the spatial information of the moving object, and the detected intensity values are equivalent to the imaging object's zero-order moment and first-order moments [30]. The object's centroid can be described by the following formula:

$$x_c = m_{10} / m_{00} = I_2 / I_1 \tag{9}$$

$$y_c = m_{01} / m_{00} = I_3 / I_1. \tag{10}$$

The positioning can be achieved by applying these three modulation patterns with three detected intensities. For the SPI system based on the DMD modulator, because each micromirror on the DMD can reflect light in the direction $+12°$, $-12°$, to fully use the advantages of DMD high-speed binary modulation, the gray geometric moment modulation matrix $S_1$, $S_2$, $S_3$ is converted into the binary matrix $S_1'$, $S_2'$, $S_3'$ by Floyd-Steinberg dithering [32], as shown in Fig. 2.(a). In contrast with hundreds of samplings to get a position [26~28], this method dramatically improves the positioning frame rate. And the sampling number is fixed as three for each positioning. The positioning accuracy is independent of the sampling number. This feature provides the condition for releasing the above constraints among the number of samples, positioning accuracy, and anti-motion blur effect. The position information $(x_c, y_c)$ of the moving object at each moment is obtained by applying geometric moment modulation, and the displacement of the moving object relative to the center of view $(x_0, y_0)$ is:

$$r_i = (r_x, r_y) = (Cx_c - x_0, Ry_c - y_0). \tag{11}$$

The above displacement $r_x$ and $r_y$ are combined with Eq.(4) to realize the motion compensation and anti-motion blur imaging. In this paper, the modulation patterns are divided into two parts: the geometric moment patterns perform the positioning, and the Hadamard patterns are used to encode and restore the image. The flow chart of the fast-location and imaging method for the moving object is shown in Fig. 2. Where Fig. 2.(a) shows the geometric moment patterns after Floyd-Steinberg dithering [32]. Then Fig. 2.(b) shows the generation of time-division multiplexing modulation patterns sequence, and the patterns are projected on the moving object shown in Fig. 2.(c). The detected signals carried different information at different times. $B$ represents the signals obtained by projecting the Hadamard patterns to the object, and

$I$ represents the signals obtained by projecting the geometric moment patterns to the object, which correspond to Eq.(1) and Eq.(7). $B$ carried the image information, and $I$ carried the position information accordingly. Fig. 2.(d) shows the image reconstruction of MCSPI. Where $S$ and $I$ are combined to obtain centroids of the object in real-time. $H$ and $B$ just yield the blurry image but combined with the centroids obtained from $S$ and $I$, the anti-motion blur image can be constructed by MCSPI.

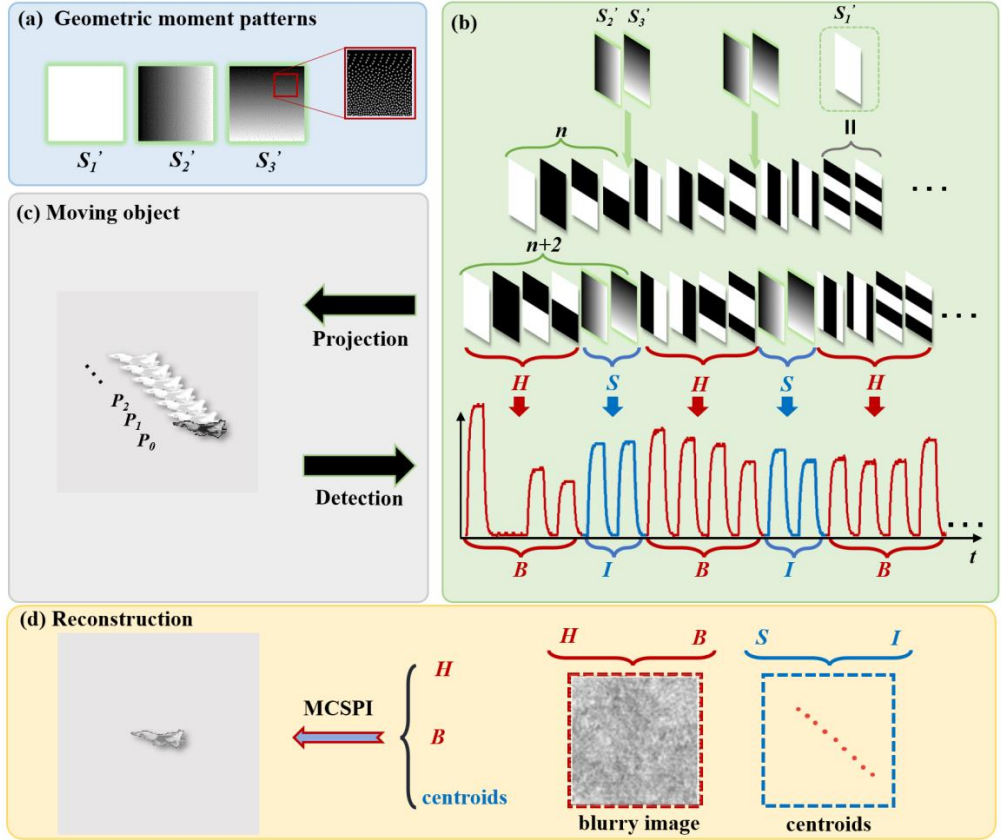

**Fig. 2.** The motion compensation single-pixel imaging method. (a) shows geometric moment patterns. (b) shows the modulation and acquisition using time multiplexing patterns sequence. (c) is a moving object. (d) is the image reconstruction process. MCSPI represents the proposed motion compensation single-pixel imaging method, $B$ and $I$ represent the detected bucket signals from Hadamard patterns and geometric moment patterns, respectively.

In the implementation of this method, to achieve high-frequency localization and anti-blur imaging of the moving object, the time multiplexing method is used. The geometric moment patterns are periodically inserted into the Hadamard patterns. As shown in Fig. 2(b), We insert two geometric moment patterns for every $n$ Hadamard patterns. Not inserting $S_1$ is that we complementary project Hadamard structured patterns with positive and negative, so the sum of the detected values of two complementary Hadamard modulations can be equivalent to the values obtained under $S_1$ illumination. $n$ can be set as different values to achieve different positioning frequencies. The system's positioning frequency $f$ of the moving object is:

$$f = 1/(n+2)\phi = \omega/(n+2), \qquad (12)$$

where $n \geq 2$, $\omega$ is the highest projection frequency of the DMD modulator. $\phi = 1/\omega$ is the minimum time for DMD to project one pattern. Every n+2 patterns is a set, including n

Hadamard and two geometric moment patterns. For each set of patterns projected, we perform one localization. Using this scheme at least four times modulation can realize one positioning and one imaging modulation. Inserting the geometric moment patterns into the Hadamard patterns sequence will reduce the imaging efficiency. Assuming that all patterns are Hadamard patterns, the imaging efficiency is 100%. The imaging efficiency of our method *Ie* is:

$$Ie = n / (n+2). \tag{13}$$

For example, when $n = 2$, compared with the traditional single-pixel imaging method, the imaging efficiency is 50%, but the positioning frequency is highest. The increase of *n* will improve the imaging efficiency, but the localization frequency will decrease accordingly. In the application, for the moving object at different speeds, the value of *n* can be flexibly set to match the optimal positioning frequency and imaging efficiency.

This MCSPI method achieves anti-motion blur imaging of the moving object based on high-frequency localization and motion compensation algorithm. Compared with the existing SPI methods [24-28] of the moving object, this method can achieve one positioning and image modulation with at least four samplings, which improves the anti-motion blur imaging performance of the moving object. Using the time-division multiplexing method and motion compensation algorithm can achieve higher anti-motion blur imaging performance for the moving object. The following experiments verify the effectiveness of this method.

## 3. Simulations

Firstly, the proposed method is verified by simulation. In the simulation, two groups of different object motions are simulated: in the first group, the object moves along a preset trajectory, with a displacement of one pixel per simulation frame; in the second group, the object's movement direction and the displacement in each frame are random. The two sets of simulation results comprehensively verify the effectiveness of the proposed method for localization and anti-motion blur imaging of the moving object.

*3.1 Moves along a preset trajectory*

In the first simulation, the object moves along the preset trajectory, a pixel for each simulation frame. The image of the simulated object is shown in Fig. 3(a). The image comes from the open-source database CIFAR10, and the resolution is 256×256. The Hadamard patterns of the same resolution are used to modulate and reconstruct the image of the object, and 65536 Hadamard patterns are employed according to the Cake-cut order. Each pattern is modulated twice according to the positive and negative, so the imaging modulation sequence contains 131072 patterns. The geometric moment pattern is inserted in the imaging modulation sequence at regular intervals, set *n*=2. The entire sequence contains 262144 modulations, and each frame performs four modulations and samples. A total of 65536 frames are simulated, the object moves 65536 pixels.

The simulation results are shown in Fig. 3. Compare with the results in Fig. 3.(b) and Fig. 3.(c), it can be seen that the traditional SPI without motion compensation cannot reconstruct the object image. But the proposed method can effectively achieve anti-blur imaging. Although the reconstructed image has apparent background noise, the subsequent simulations will verify that the noise will gradually weaken with increased samples. Fig. 3.(d) shows the simulation result obtained using Wu et al.'s method [26]. According to its definition, set $N^w = 30, N_1^w = N_2^w = 15$, where $N^w$ represents the total number of modulates and samples during the time interval considered in the quasi-stationary state. $N_1^w$ and $N_2^w$ are the samples used for localization and imaging, respectively. The result shows it cannot reconstruct a clear object image in the same simulation environment because of the low positioning frequency and accuracy when the object moves fast. This method is limited by the mutual constraints of the number of samples, positioning accuracy, and anti-motion blur effect. The performance of this method is difficult to continue to improve. Fig. 3.(e) shows the ground truth of the object position and the position obtained by two methods. The ground truth is the red line in the figure, and the green spots are the measurement result of our method. The average absolute error of positioning is 1.3 pixels using our method. The blue square symbols are the positions obtained using the method of Wu et al. [26]. Fig. 3.(f) and Fig. 3.(g) show the object's real positions and calculated positions in the x-direction and y-direction. It can be seen that our method can obtain accurate positioning parameters with high frequency. Anti-motion blur imaging is thus achieved.

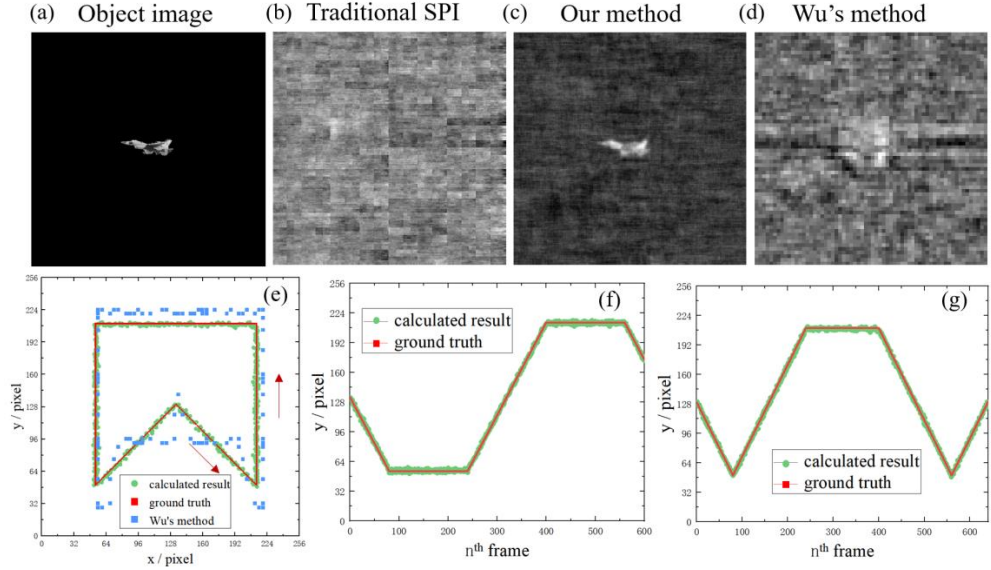

**Fig. 3.** The positioning and imaging results of the simulation: (a) is the moving object; (b) is the traditional SPI result without motion compensation; (c) is the object image reconstructed by our method; (d) is the imaging reconstructed by Wu's method [26]; (e) is the comparison between the real and calculated positions by our method; (f) and (g) are the object's real and calculated positions of our method in the *x* and *y* directions.

*3.2 Moves randomly*

The second simulation uses the resolution board as the simulated object, with the image resolution being 256×256. The object is within the simulation field and displaces different distances (0 ~ 60 pixels) along a random direction in each frame. The sampling and reconstruction process is the same as simulation 3.1. The total simulation frame is 1638400, each frame performs four modulations and samples, corresponding to the number of samples $\tau$ = 6553600. The object is modulated cyclically with the modulate sequence mentioned in simulation 3.1.

The simulation result is shown in Fig. 4. Comparing Fig. 4.(b) and Fig. 4.(c), it can be seen that this method can achieve better imaging results than the method without motion compensation. It can be seen from Fig. 4(c~f) that as the number of modulation samples increases, the background noise gradually decreases, and the image details become clearer. The results are quantitatively analyzed using the parameters MSE and PSNR, which can be expressed as :

$$MSE = \frac{1}{CR}\sum_{x=1}^{C}\sum_{y=1}^{R}\left[P'(x,y) - P(x,y)\right]^2 \qquad (14)$$

$$PSNR = 10\log_{10}(peakval^2 / MSE), \qquad (15)$$

where the image resolution is *C*×*R*, and *peakval* represents the peak grayscale value of the reconstructed image. Fig. 4 (g) shows the relationship between the peak signal-to-noise ratio (PSNR) and the mean square error (MSE) of the reconstructed image with the number of samples τ. According to the results, it can be seen that with the samples increase, the MSE gradually decreases, and the PSNR gradually increases. The PSNR reaches 20.83dB when τ = 6553600. That means that the proposed method's imaging quality will gradually improve with the time of stable object tracking.

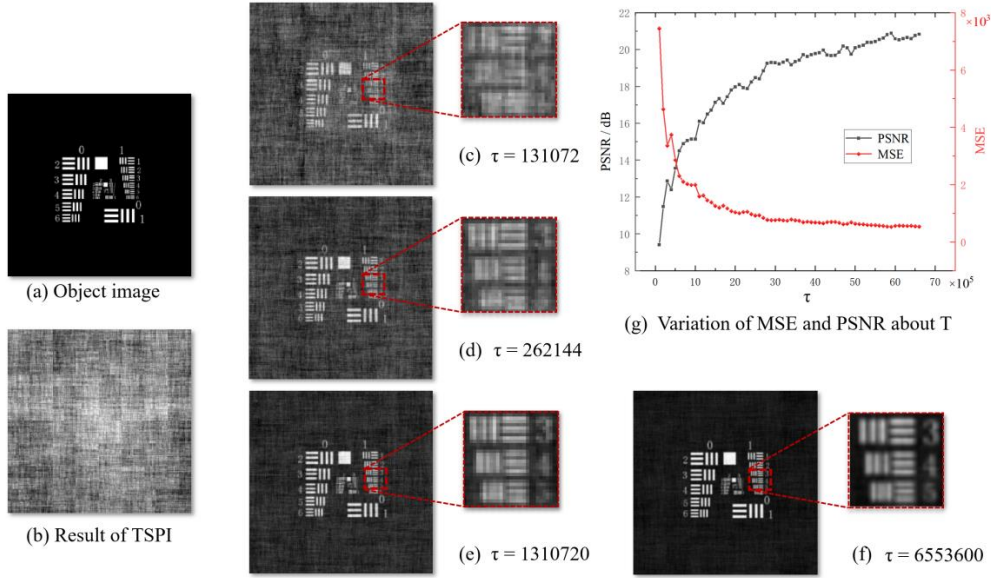

**Fig. 4.** Resolution object reconstruction results: (a) object image; (b) traditional SPI result; (c~f) object images reconstructed by the proposed method under different modulation and sampling number τ; (g) the relationship between the reconstructed image MSE and PSNR with the number of samples τ.

## 4. Experiments

In this part, we verify the proposed method experimentally. The experimental system device includes a light-emitting diode (LED) with a maximum power of 200 mW (M530L4-C1, Thorlabs), a digital micromirror device (DMD, Texas Instruments Discovery V7000 with 1024×768 micromirrors)), a lens group, a photomultiplier (PMM02, Thorlabs), a high-speed digitizer card (PicoScope 6407, PICO), a computer, and a 2-D motorized stage. The structure is shown in Fig. 5. The DMD is used to modulate the light from the light source and provide structured illumination onto a scene through a projection lens (PL). A 2-D motorized stage moves the object in two directions. Light reflected from the object is collected by the photomultiplier tube (PMT) through a collection lens (CL). The high-speed digitizer obtains the output signal from the photomultiplier. In the experiment, 512×512 micro-mirrors in the middle of the DMD were used for modulation, producing the structure light with 512×512 resolution. The high-speed digitizer was used at a speed of 200MHz, and the modulation frequency of DMD was set to 22.2kHz. The Hadamard patterns used in the experiment are 512×512, which are the first 4096 patterns of the Cake-cut order Hadamard patterns. Each pattern is projected twice according to the positive and negative, corresponding to the 8192 projections. Set $N = 2$, the positioning frequency can be obtained by Eq.(10) for 5.55kHz. An entire projection sequence includes 16384 projections with localization and imaging. The experiments are carried out with two different motions of the object.

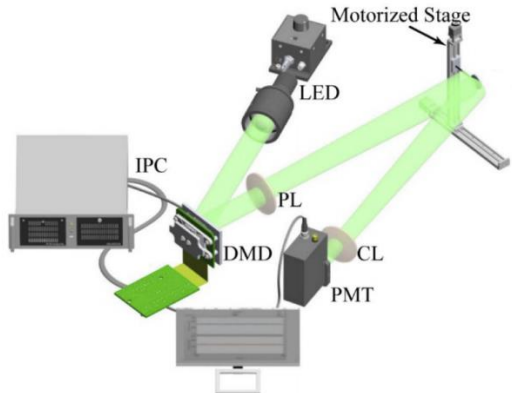

**Fig. 5.** Experimental setup. A light-emitting diode (LED) uniformly illuminates a digital micromirror device (DMD), used to provide structured illumination onto a scene through a projection lens (PL). A 2-D motorized stage moves the object in two directions. Light reflected from the object is collected by a photomultiplier tube (PMT) through a collection lens (CL). The measured light intensities are used in our reconstruction algorithm run on an industrial personal computer (IPC) to reconstruct the image of the object.

*4.1 Simple pendulum Motion*

In the first experiment, the object is fixed by a thin line and performs a pendulum motion in the illumination area. Fig. 6(a) shows the object's real trajectory and calculated position. The red line represents the real trajectory. The black points represent the calculated positions of our method. The mean absolute error is 5.1 pixels. Fig. 6(b) is the single-pixel imaging result without motion compensation. There is significant imaging motion blur. Fig. 6(c) is the single-pixel imaging result of the object in a static state. Compared with Fig. 6(d), it can be seen that when sampling number $\tau = 16384$, the obtained image can distinguish the object, but there is apparent background noise. Fig. 6(e) shows that a clear object image can be obtained under the modulating and sampling number $\tau = 163840$, with low background noise and high contrast. The motion video of the object can also be reconstructed by our method (see Visualization 1). The pseudo resolution of the image is 512×512 when the object is moving. Comparing Fig. 6(d) and (e), it can be seen that the background noise can be effectively suppressed under the higher sampling number, and the image with a higher signal-to-noise ratio can be obtained.

*4.2 Moves with the motorized stage*

In the second experiment, the motorized stage moves the object according to the set trajectory, and the peak angular velocity relative to the imaging system is about 0.5rad/s. We use the same location and reconstruction method as Experiment 4.1. The results are shown in Fig. 7. The first row shows the imaging results of the moving object without motion compensation. The apparent motion blur appears in the reconstructed image. Three blurred object contours were reconstructed at the slow-moving turns. The second row of Fig. 7(a) is our method's motion compensation single-pixel imaging result. Under the low sampling number, the reconstructed image has apparent background noise. However, with the increase of the sampling number, the reconstruction quality of the object image gradually improves, and the background noise gradually weakens. Fig. 7(b) and (c) are the object positions calculated by our method. Fig. 7(d) shows that the restored image's peak signal-to-noise ratio (PSNR) gradually increases as the number of samples increases. The reference image used in the calculation of *PSNR* is the imaging result of the object in a stationary state. When the number of samples τ = 163840, the *PSNR* of the restored image is 17.3dB. The motion video of the object can be reconstructed by our method (see Visualization 2).

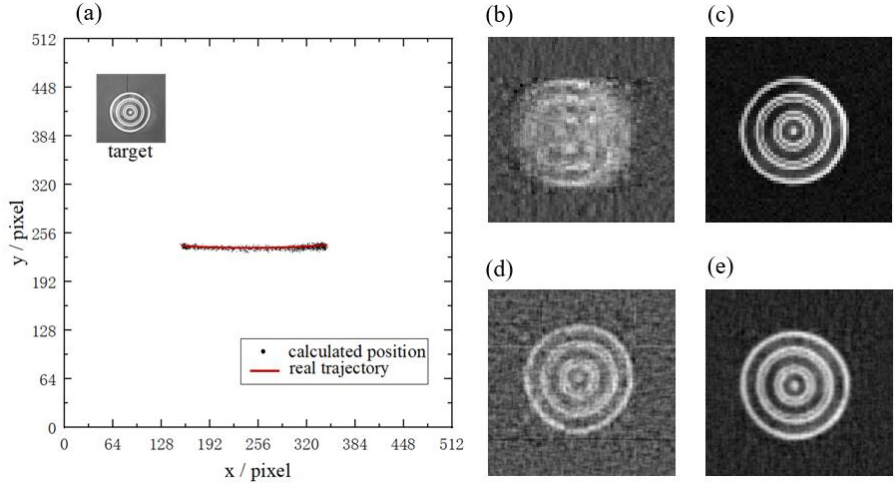

**Fig. 6.** Reconstruction results of a pendulum moving object: (a) object's real trajectory and calculated positions; (b) result without motion compensation; (c) result of the static object; (d) result of our method with modulation and sampling number τ = 16384; (e) result of our method with modulation and sampling number τ = 163840.

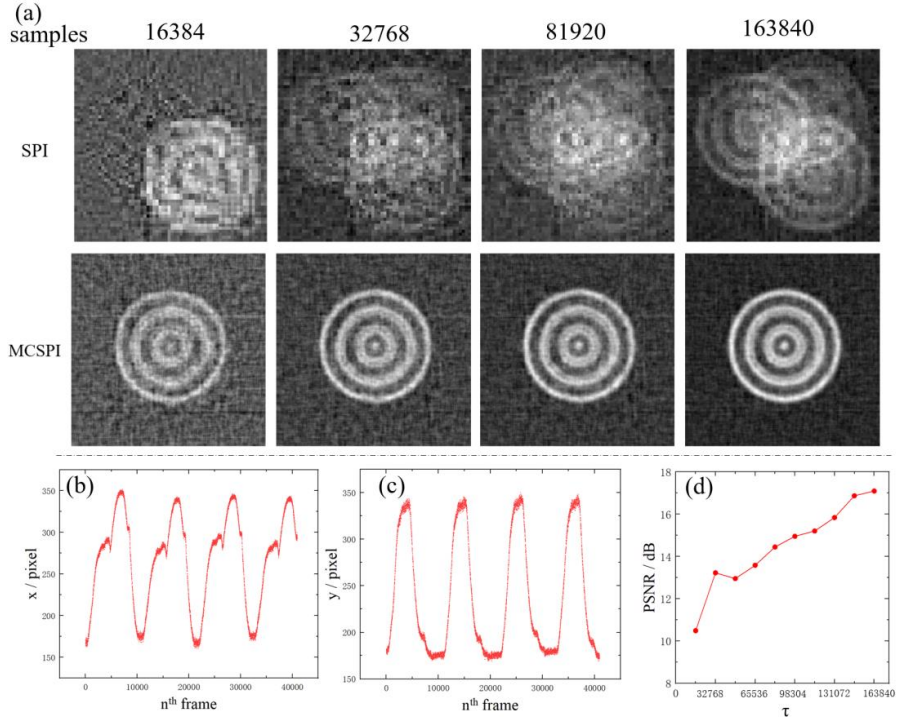

**Fig. 7.** Imaging and positioning results of the moving object with the motorized stage: (a) the first row is the reconstruction results without motion compensation under different projection sampling numbers; the second row is the imaging results using our method under different projection and sampling numbers. (b) shows the positions measured by our method in the x-direction. (c) shows the positions measured by our method in the y-direction. (d) is the variation of *PSNR* with the number of samples τ.

In our experiment, the reconstruction of images is performed in off-line approach. We save the collected bucket detection signals, and then transmit the data to the memory and processor of the computer. Using our algorithm to iteratively reconstruct the image of the object. In this experiment, every four patterns is a set, including two Hadamard patterns and two geometric moment patterns. For each set of patterns projected, it corresponds to 4 measurements. Based on these measurements, our algorithm performs once object localization and image iteration. The image quality gradually higher as the number of samples increases and the number of iterations increases. As shown in the second row of Fig.7(a).

The modulation frequency of our system is 22.2kHz. The projection and sampling times used for the images in the second row of Fig.7 can be calculated as about 0.74s, 1.48s, 3.69s, and 7.38s, respectively. Since each positioning and image iteration requires four pattern projections that take only 180us, a video of moving objects can be constructed by combining object positions and the continuously updated object image, such as Visualization 1 and Visualization 2.

In the experiment, $n = 2$ is set, and the positioning frequency is 5.55kHz. The above results show that the proposed MCSPI method has the anti-motion blur imaging ability. Compared with the existing single-pixel imaging methods [26-28] for moving objects, our method ensures clear imaging of the fast-moving object based on the motion compensation algorithm and more real-time positioning performance. This method meets the needs of single-pixel imaging in more motion scenes and broadens the scope of application of single-pixel imaging.

## 5. Conclusion

In this paper, we proposed a motion compensation single-pixel imaging method used time-division multiplexing. This method uses the geometric moment patterns and Hadamard patterns to alternately encode the object's position information and image information with time-division multiplexing. Then, we used the motion compensation reconstruction algorithm to achieve anti-blur imaging of the moving object. Because we position the moving object with a high frame rate, this method improves the anti-motion blur single-pixel imaging performance and meets the needs of single-pixel imaging in more motion scenes. This method does not require additional hardware to locate or estimate the motion state of the object. In the scene where the spatial structure of the object is unchanged and the background is static, it has the ability of stable tracking and clear imaging, which provides a new solution for single-pixel imaging of the moving object. It has application value in low-light, hyperspectral, long-distance moving object imaging, and other similar fields [33-35]. However, this method also has limitations at present. For example, if the object rotates, the proposed method will produce rotate motion blur. In addition, when there are multiple objects in the scene or imaging in a complex background, the localization will also be affected, and the quality of the reconstructed image will also be reduced. Moreover, restoring high-quality images requires long-term accumulation, which is not conducive to dynamic real-time imaging. Therefore, improving the algorithm to achieve real-time high-definition imaging is necessary. These questions are the focus of our next research.

*CRediT authorship contribution statement*

*Declaration of Competing Interest*

The authors declare that they have no known competing financial interests or personal relationships that could have appeared to influence the work reported in this paper.